\documentclass[12pt,preprint]{aastex}

\newcommand{\lta}{{\>\rlap{\raise2pt\hbox{$<$}}\lower3pt\hbox{$\sim$}\>}}
\newcommand{\gta}{{\>\rlap{\raise2pt\hbox{$>$}}\lower3pt\hbox{$\sim$}\>}}

\begin{document}

\title{THE AGE DISTRIBUTION OF MASSIVE STAR CLUSTERS 
       IN THE ANTENNAE GALAXIES}

\author{S. Michael Fall, Rupali Chandar, and Bradley C. Whitmore}
\affil{Space Telescope Science Institute, 3700 San Martin Drive, 
Baltimore, MD 21218}
\affil{fall@stsci.edu, rupali@stsci.edu, whitmore@stsci.edu}

\begin{abstract}

We determine the age distribution of star clusters in the 
Antennae galaxies (NGC~4038/9) for two mass-limited samples 
($M > 3\times10^4$~$M_{\odot}$ and $M > 2\times10^5$~$M_{\odot}$).  
This is based on integrated broadband $UBVI$ and narrowband 
$H\alpha$ photometry from deep images taken with the 
{\it Hubble Space Telescope}. 
We find that the age distribution of the clusters declines 
steeply, approximately as $dN/d\tau \propto \tau^{-1}$. 
The median age of the clusters is $\sim$$10^7$~yr, which we 
interpret as evidence for rapid disruption (``infant mortality'').
It is very likely that most of the young clusters are not 
gravitationally bound and were disrupted near the times they 
formed by the energy and momentum input from young stars to the 
interstellar matter of the protoclusters.  
At least 20\% and possibly all stars form in clusters and/or 
associations, including those that are unbound and short-lived. 

\end{abstract}



\section{INTRODUCTION}

The age distribution of a population of star clusters contains
valuable information about the formation and disruption of the
clusters.
In this Letter, we present the age distribution of massive
star clusters in the Antennae galaxies (NGC~4038/9), derived 
from multicolor images taken with the Wide Field Planetary 
Camera 2 (WFPC2) on the {\it Hubble Space Telescope (HST)}.
We have previously used these observations to determine the
luminosity function, mass function, and space distribution
of the clusters (Whitmore et al. 1999; Zhang \& Fall 1999; 
Zhang, Fall, \& Whitmore 2001). 
See Fall (2004) for an overview of this work and an early
presentation of the age distribution.

The star clusters in the Antennae galaxies have attracted 
attention for several reasons.
The Antennae are the nearest and best-studied pair of merging 
galaxies, consisting of two large spirals that collided and 
began to commingle a ${\rm few} \times 10^8$~yr ago.
The number of young clusters in this system is huge, permitting 
the mass, age, and space distributions to be determined better 
than in any other galaxies, except perhaps the Milky Way and 
Andromeda.
The ongoing merger is almost certainly responsible in some way 
for this large population of clusters.
Understanding the formation and disruption of clusters in this
setting is important because it represents a latter-day example
of the hierarchical formation of galaxies, a process that 
operated even more effectively in the early universe.

\section{OBSERVATIONS AND AGE ESTIMATES}

Our {\it HST} observations of the Antennae galaxies are described 
fully by Whitmore et al. (1999).
The images were taken in 1996 January with the WFPC2 for 
2000--4500~s with each of the broadband filters F336W ($U$), 
F439W ($B$), F555W ($V$), and F814W ($I$), and for 3800~s with 
the narrowband F658N ($H\alpha$) filter.
Point-like objects (stars and clusters) were identified and their 
total magnitudes were measured with the IRAF task DAOPHOT.  
The photometry in this instrumental system was then converted to 
the Johnson-Cousins $UBVI$ system using the color transformations 
from Holtzman et al. (1995).
Approximately 11,000 objects were detected with $17.4 < V < 25.4$, 
corresponding to $-14 < M_V < -6$ at the adopted distance of 
19.2 Mpc.

We estimate the age $\tau$ and extinction $A_V$ of each 
cluster by comparing the observed magnitudes in the five bands 
with those from stellar population models. 
From $A_V$ and the observed $V$ magnitude of a cluster, we then 
determine its extinction-corrected luminosity $L_V$ and position
in the $L_V$-$\tau$ plane.
Specifically, we use the Bruzual \& Charlot (2003) models with 
solar metallicity and Salpeter initial mass function, and we 
assume Galactic-type extinction (Fitzpatrick 1999).  
The best-fit values of $\tau$ and $A_V$ are those that minimize
the statistic
$$
\chi^2(\tau,A_V) = \sum_{\lambda}
W_{\lambda}~(m_{\lambda}^{\mbox{obs}} 
- m_{\lambda}^{\mbox{mod}})^2,
$$
where $m_{\lambda}^{\mbox{obs}}$ and $m_{\lambda}^{\mbox{mod}}$ 
are the observed and model magnitudes, respectively, and the sum 
runs over all five bands, $\lambda=U,B,V,I$, and $H\alpha$.
For the $H\alpha$ band, we include the total flux---line 
plus nearby continuum---in both $m_{\lambda}^{\mbox{obs}}$ and 
$m_{\lambda}^{\mbox{mod}}$.
This procedure avoids the large uncertainties inherent in the 
continuum-subtracted fluxes of clusters with little or no line 
emission (i.e., the differences between two nearly equal but 
imperfect measurements). 
The weight factors in the formula for $\chi^2$ are taken to 
be $W_{\lambda} = [(0.05)^2 + \sigma_{\lambda}^2]^{-1}$, where
$\sigma_{\lambda}$ is the photon noise (in magnitudes) for each 
band, and the additional 0.05 mag of uncertainty was determined 
by false-object measurements in the same {\it HST} images 
(Whitmore et al. 1999).  

Some of the ionizing photons produced by the clusters will 
escape from their immediate vicinity and cause H$\alpha$ 
emission at locations outside our measurement aperture 
of radius 0.3''.
We correct for this, in an average sense, by reducing 
the number of ionizing photons in the models by an escape
fraction $f_e$.
The value of $f_e$ is determined by matching the mean
predicted and observed H$\alpha$ emission of the youngest 
clusters.
In this way, and assuming the usual case B recombination rate, 
we estimate $f_e \approx 0.4$.
This is similar to the value of $f_e$ found in studies of HII 
regions in the Milky Way and other galaxies in the Local Group 
(Terebey et al. 2003 and references therein).

We assess the reliability of our age-fitting procedure as 
follows.
From $\chi^2$, we compute an internal 1$\sigma$ error of 
0.2 in $\log\tau$.
The external 1$\sigma$ error is larger, 0.3--0.4 in
$\log\tau$, corresponding to a factor of 2.0--2.5 in $\tau$.
We determine this level of accuracy by comparing the ages of
10 young clusters in the Antennae galaxies derived from our 
integrated $UBVIH\alpha$ photometry with independent ages 
of the same clusters derived from UV spectra taken with the 
Goddard High Resolution Spectrograph and the Space Telescope 
Imaging Spectrograph on {\it HST} (Whitmore 2001; R.~Chandar 
et al. 2005, in preparation).
We note that a similar external 1$\sigma$ error has been 
determined by comparing the ages of nearby clusters (in the 
LMC) derived from both integrated colors and main-sequence 
turnoffs (Elson \& Fall 1988).

A more serious concern here is that the errors in some of 
our photometric ages have a non-random component.
The integrated light from massive clusters with $8 \times10^6 
< \tau < 1.6 \times10^7$~yr (and high metallicity) is dominated 
by red supergiant (RSG) stars (Leitherer et al. 1999; Bruzual
\& Charlot 2003).
During this phase, the integrated colors change so abruptly that 
the fitted ages, in the presence of observational errors, become
degenerate, with a strong tendency to avoid values just above
$10^7$~yr.
Fortunately, these non-random errors are also relatively small, 
0.4 or less in $\log\tau$, and they therefore have little effect 
on the overall shape of the age distribution when examined on 
scales of 0.8 or more in $\log\tau$.

Some of the brightest clusters are spatially resolved in the 
{\it HST} images, but those near the limiting magnitude are 
indistinguishable from stars.  
We minimize stellar contamination in our sample of clusters by 
restricting it to objects brighter than all but the most luminous 
stars ($L_V > 3 \times 10^5 L_{\odot}$).  
Since our sample is optically selected, it undoubtedly excludes 
some clusters that are heavily obscured by dust.  
However, from a comparison of the locations of radio-continuum 
and optical sources in the Antennae galaxies, Whitmore \& Zhang 
(2002) estimate that the present sample of clusters is 
$\sim$$85$\% complete.

\section{AGE DISTRIBUTION}

Figure~1 shows the resulting luminosity-age distribution of 
the clusters. 
This is an updated version of a similar diagram based on two 
reddening-free $Q$ parameters derived from our $UBVI$ photometry
(Zhang \& Fall 1999). 
The new diagram, which is based on the $H\alpha$ as well as 
the $UBVI$ measurements, helps to distinguish clusters younger
and older than $\tau \approx 6 \times 10^6$~yr, the age at which 
the ionizing flux from a stellar population declines rapidly. 
The vertical gap at $\tau \approx (1 \rightarrow 2) \times 
10^7$~yr in Figure~1 is an artifact caused by the rapid evolution 
of the integrated colors of clusters during the RSG phase, as 
discussed above.
This small-scale feature, while visually prominent, has a 
negligible bearing on our conclusions. 
The diagonal lines in Figure~1 represent the evolutionary tracks 
of stellar population models of fixed initial mass, $M = 3\times
10^4$~$M_{\odot}$ and $2\times10^5$~$M_{\odot}$.
The horizontal line at $L_V = 3\times10^5$~$L_{\odot}$ is
our adopted limit for stellar contamination.

Figure~1 contains much of the available statistical information 
about the population of star clusters in the Antennae galaxies.
The luminosity function, $\phi(L) \equiv dN/dL$, for example, is
obtained by projecting the two-dimensional distribution horizontally
along the age axis.
The mass function, $\psi(M) \equiv dN/dM$, is obtained by projecting 
instead in a diagonal direction, along the fading tracks, 
and counting clusters in the corresponding mass bins.
This procedure yields a mass function for the young clusters 
($\tau < 10^8$~yr) that can be approximated by a power law of the 
form $\psi(M) \propto M^{\beta}$, with $\beta \approx -2$ over the 
observed range of masses, $10^4 < M < 10^6$~$M_{\odot}$ (see Fig.~2
of Fall 2004), thus confirming the earlier result of Zhang \& 
Fall (1999).
 
The age distribution can be derived from Figure~1 in different ways, 
depending on the selection criteria of the sample.
The form of $dN/d\tau$ usually presented in the literature is for a 
luminosity-limited sample, obtained simply by counting clusters in 
age bins above the limiting luminosity.
This, however, is not straightforward to interpret in dynamical
terms because it depends on the fading of the clusters by stellar
evolution in addition to their formation and disruption histories.
A more physically informative age distribution is that for a
mass-limited sample.
We obtain this form of $dN/d\tau$ by counting the clusters in age 
bins above the diagonal fading tracks in Figure~1.  

Figure~2 shows the age distribution for both luminosity- and 
mass-limited samples of clusters.\footnote{
These age distributions include completeness corrections 
from Whitmore et al. (1999), but they are very similar to the
corresponding distributions without corrections over the ranges 
of masses and ages plotted here.}
The former is significantly steeper than the latter because 
it includes a higher proportion of young clusters and a lower 
proportion of old clusters.
In the following, we only consider results from the mass-limited 
samples.
We find that the age distribution declines steeply, starting at 
very young ages, and has no obvious bend or other features.
It can be approximated by a power law of the form $dN/d\tau 
\propto \tau^{-1}$, corresponding to $dN/d{\log\tau} = 
{\rm const}$, over the range $10^6 < \tau < 10^9$~yr, with 
little, if any, dependence on mass, at least for $M > 3 \times 
10^4$~$M_{\odot}$.
This general behavior is also apparent in the $L_V$-$\tau$
diagram (Fig.~1); there are roughly as many clusters above
the diagonal fading tracks with ages in the range $10^6 < \tau 
< 10^7$~yr as there are in the ranges $10^7 < \tau < 10^8$~yr 
and $10^8 < \tau < 10^9$~yr.

Another useful way to characterize the age distribution is in
terms of the median age $\tau_m$ of the clusters. 
For the mass-limited samples, ranking of the fitted ages gives 
$\tau_m \approx 2 \times 10^7$~yr.
We suspect that the true median age is smaller than this, for 
three reasons: 
(1) the RSG gap in Figure~1 artificially excludes individual
ages and hence values of $\tau_m$ in the range $(1 \rightarrow 2) 
\times 10^7$~yr; 
(2) nearly half of the clusters in the mass-limited samples
have strong H$\alpha$ emission, indicating that they are
younger than $6 \times 10^6$~yr; and
(3) incompleteness in our sample caused by dust obscuration is 
expected to be more severe for young clusters than for old ones.
We therefore adopt, as a rough estimate, $\tau_m \sim 10^7$~yr.

We have performed a variety of tests to determine the reliability
of these results.
We have repeated the entire analysis with ages derived with and
without extinction corrections, with and without the $H\alpha$
measurements, and with both the Bruzual \& Charlot (2003) and 
Leitherer et al. (1999) stellar population models.
In all cases, the age distribution is very similar to that shown
in Figure~2. 
We have also performed Monte Carlo simulations in which the 
input age distribution was specified, and the integrated colors 
of the clusters were computed from stellar population models
and perturbed by random observational errors.
The ages of the simulated clusters were then estimated by 
minimizing $\chi^2$ and the resulting output age distribution 
was compared with the input distribution.
This again confirms the results shown in Figure~2.
We conclude that the steep decline in the age distribution 
is robust relative to any likely errors in our analysis.

\section{INTERPRETATION}

The age distribution of a population of star clusters in general 
reflects a combination of their birth and death rates.
Hence, the steep decline of $dN/d\tau$ in the Antennae galaxies
might be interpreted as the result of a brief but intense burst 
of cluster formation, possibly triggered by the current interaction 
between the two galaxies. 
This, however, seems very unlikely for two reasons.
First, the Antennae galaxies have been interacting for the past 
${\rm few} \times 10^8$~yr, a time much longer than the median age
of the clusters, $\sim$$10^7$~yr. 
Second, the age distribution is similar in different parts 
of the galaxies, some of which appear to be interacting more
strongly than others (see Fig.~3 of Whitmore 2004).
It is hard to understand how a burst of cluster formation
could be synchronized so precisely over such large distances.
Thus, we prefer to interpret the steep decline in the age
distribution in terms of the disruption of the clusters.

The short timescale on which the clusters are disrupted 
indicates that most of them are not gravitationally bound.
In terms of the initial characteristic radius $R_0$ 
(three-dimensional median radius) and virial velocity 
$V_0$, the crossing time for stars orbiting within a bound 
cluster or protocluster is $\tau_{\rm cr} = R_0/V_0$. 
We estimate $\tau_{\rm cr} \approx 10^6$~yr for a typical 
young cluster with $M = 10^5$~$M_{\odot}$, $R_0 = 5$~pc, 
and $V_0 = (0.4GM/R_0)^{1/2} = 6$~km~s$^{-1}$.
The crossing time is expected to be similar for clusters of 
different masses and radii because it depends on these quantities
only through the mean density, which is determined primarily by 
the tidal field of the host galaxy.
If a protocluster suddenly lost most of its mass by the removal 
of interstellar matter (ISM), it would no longer be gravitationally 
bound and would expand almost freely, its characteristic radius 
increasing with age as $R(\tau) \approx R_0 (\tau/\tau_{\rm cr})$ 
and its characteristic surface density decreasing as $\Sigma(\tau) 
\approx \Sigma_0(\tau/\tau_{\rm cr})^{-2}$.
Thus, after $\tau \sim 10 \tau_{\rm cr} \sim 10^7$~yr,
the surface brightness of the cluster (even ignoring the fading
by stellar evolution) would be roughly a factor of $10^2$ lower
or 5 mag fainter than initially (at $\tau \sim 10^6$~yr), 
and it would then disappear among the statistical fluctuations 
in the foreground and background of field stars.\footnote
{The expansion of the clusters may be difficult to verify 
observationally because it is unlikely to be homologous; 
the more prominent inner cores of the clusters may expand 
less rapidly than the outer envelopes or may even contract 
as a result of stellar mass segregation. Moreover, even with 
{\it HST}, only the brightest clusters appear resolved.}

What could cause this high rate of ``infant mortality''?
The gravitational binding energy of a massive cluster 
is only $\sim$$10^{50}$~ergs, much less than the energy 
produced by a single massive star over its short lifetime. 
The energy and momentum output from massive stars comes in the 
form of ionizing radiation, stellar winds, jets, and supernovae. 
These processes could easily remove much of the ISM
from a protocluster, leaving the stars within it 
gravitationally unbound and expanding freely as argued above, 
even if the cloud in which they formed was initially bound.
This is the standard explanation for the expanding OB
associations in the Milky Way (Hills 1980).
It is also possible that parts of the protoclusters or their
parent molecular clouds were never gravitationally bound. 
We propose that the same processes account for the steep age 
distribution of the massive clusters in the Antennae galaxies.
It is worth noting here that these clusters are often regarded 
as young globular clusters.

We can understand this result as follows.
The energy and momentum input to a protocluster are approximately
proportional to the number of massive stars within it and hence
to its mass; a protocluster with more ISM to remove has more 
stars to do it.
Consequently, we expect the fraction of disrupted clusters 
to be roughly independent of their masses.
This is consistent with our observations that the shape of the 
mass function is nearly independent of age, at least for $\tau 
< 10^8$~yr, and that the shape of the age distribution is nearly 
independent of mass, at least for $M > 3 \times10^4$~$M_{\odot}$.
Moreover, because the clusters are disrupted mainly by internal 
processes, at least initially, we expect the age distribution 
to be largely independent of the properties of the host galaxy.
This prediction could be tested by comparing the cluster 
populations in different galaxies.
In fact, the age distribution presented here for the clusters 
in the Antennae galaxies is similar to that for clusters in 
the solar neighborhood, albeit with very different masses 
(Lada \& Lada 2003).

These arguments suggest that the survival of a cluster, not 
its disruption, may be the more difficult fact to explain.
Whether a particular cluster survives may depend on 
``accidental'' factors, such as just where and when the 
most massive stars happen to form within the protocluster.
Indeed, the inner, dense cores of protoclusters are more likely
to survive than their outer envelopes.
As a result, clusters may retain some of their stars and 
lose others.
Because the mass function of the clusters is a power law, roughly 
independent of age, we cannot distinguish between the case in which 
every cluster loses half of its mass in the first $\sim$$10^7$~yr
and the case in which half of the clusters lose all of their mass 
while the others lose none---or any other case between these extremes.
However, we can conclude, irrespective of this ambiguity, that half
of the stars that form in recognizable clusters are dispersed in 
the field population before they are $\sim$$10^7$~yr old. 

In this connection, it is also interesting to estimate the 
fraction of stars that are born within clusters and/or associations.
About 20\% of the total H$\alpha$ emission in the Antennae 
galaxies occurs at the locations of the clusters in our sample.
This is a lower limit on the fraction of stars born in clusters,
for several reasons: (1) it includes only those clusters brighter
than our stellar contamination limit and hence more massive than 
$\sim$$10^4$~$M_{\odot}$; (2) some of the ionizing radiation from 
the clusters will escape from their immediate vicinity and cause
H$\alpha$ emission elsewhere; and (3) some clusters will be 
disrupted even before they stop producing ionizing radiation. 
Each of these effects alone could increase the fraction by a factor
of 1--2, and together they could increase it by a factor up to 5.
We conclude from this that at least 20\% and possibly all stars 
were born in clusters and/or associations, most of which are
disrupted rapidly. 

The few clusters that manage to survive their infancy are 
subject to disruption on longer timescales by a variety of 
stellar dynamical processes, including dynamical friction, 
internal two-body relaxation, and external gravitational shocks.
Of these, two-body relaxation is by far the dominant mechanism 
for low-mass clusters over long times ($\tau \ga {\rm few} 
\times 10^8$~yr). 
This causes the mass function of the clusters, although initially 
a power law, to evolve in a Hubble time into one with a peak or 
turnover at $M_p \approx 2 \times 10^5$~$M_{\odot}$, similar to 
that of old globular clusters (Fall \& Zhang 2001 and references 
therein).
Since, as shown here, most clusters are disrupted more rapidly
than this, bound, long-lived clusters account for only a tiny
fraction of the total stellar population of a galaxy
($10^{-4}$ to $10^{-3}$ by mass). 

\section{CONCLUSIONS}

We have used integrated broadband $UBVI$ and narrowband 
$H\alpha$ photometry from deep {\it HST} images to determine 
the age distribution of star clusters in the Antennae galaxies 
for two mass-limited samples ($M > 3\times10^4$~$M_{\odot}$
and $M > 2\times10^5$~$M_{\odot}$). 
From this, we draw the following conclusions:

1. The age distribution of the clusters declines steeply,
approximately as $dN/d\tau \propto \tau^{-1}$. 

2. The median age of the clusters is $\sim$$10^7$~yr,
which we interpret as evidence for rapid disruption 
(``infant mortality'').  

3. It is very likely that most of the young clusters are not 
gravitationally bound, and were disrupted near the times they 
formed by the energy and momentum input from young stars to 
the ISM of the protoclusters.  

4. At least 20\% and possibly all stars form in clusters and/or 
associations, including those that are unbound and short-lived.

\acknowledgments

We thank Claus Leitherer, Jes\'us Ma\'{\i}z-Apell\'aniz, and
the referee for helpful comments.
This work was partially supported by NASA grants GO-05416.01-93A 
and GO-07468.01-A.
{\it HST} data are obtained at STScI, which is operated by AURA,
Inc., under NASA contract NAS5-26555.

\clearpage

\begin{figure}
\plotone{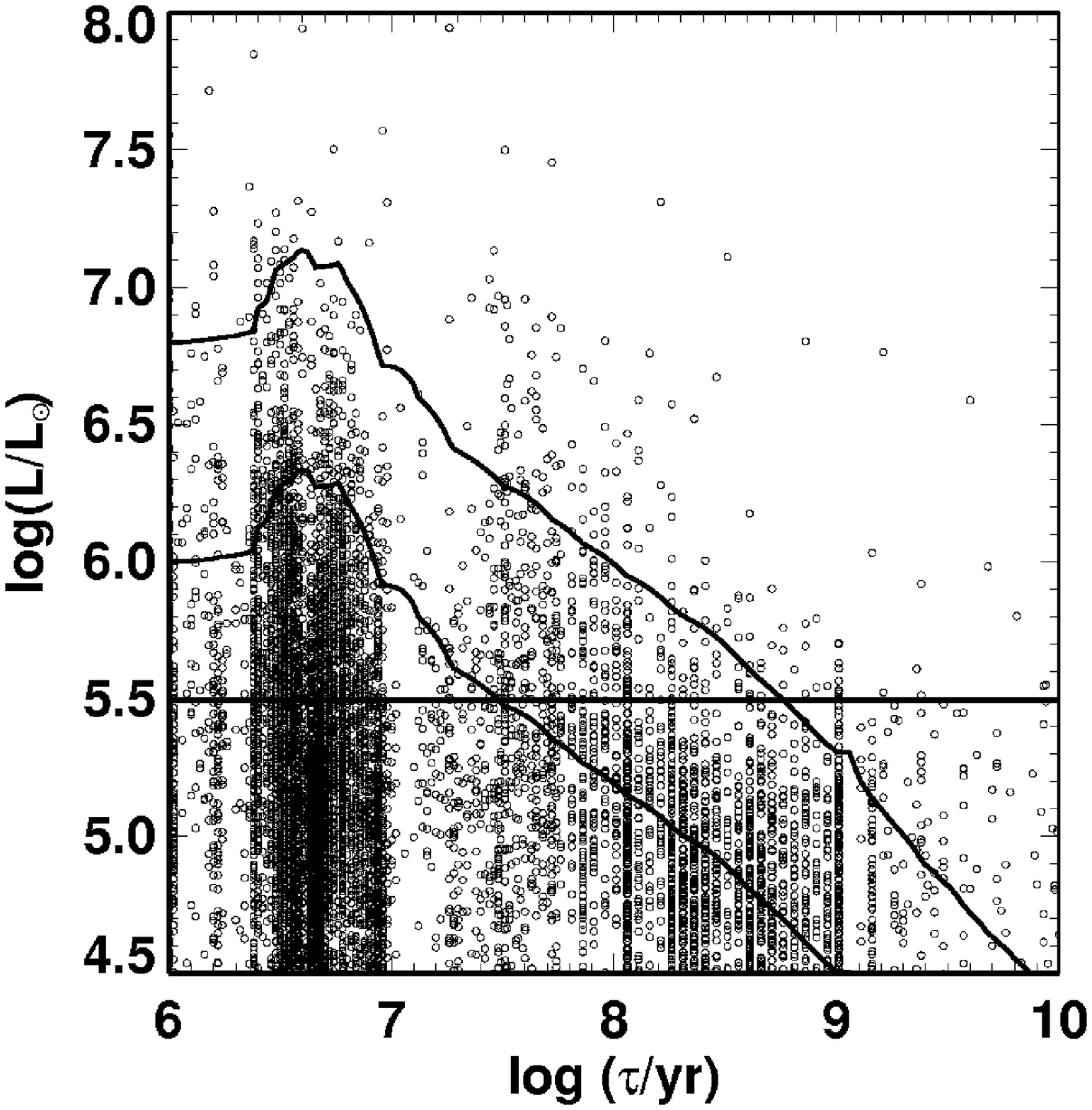}
\caption{Luminosity-age distribution of star clusters in the
Antennae galaxies.
$L$ is the extinction-corrected luminosity in the $V$ band. 
The diagonal lines represent the evolutionary tracks of stellar 
population models with initial masses of $3\times10^4$~$M_{\odot}$ 
and $2\times10^5$~$M_{\odot}$ (Bruzual \& Charlot 2003). 
The horizontal line at $L = 3\times10^5$~$L_{\odot}$ shows the 
approximate upper limit for stellar contamination.
The vertical gap at $\tau \approx (1 \rightarrow 2) \times 
10^7$~yr is an artifact of the age-fitting procedure (see text).}
\end{figure}

\begin{figure}
\plotone{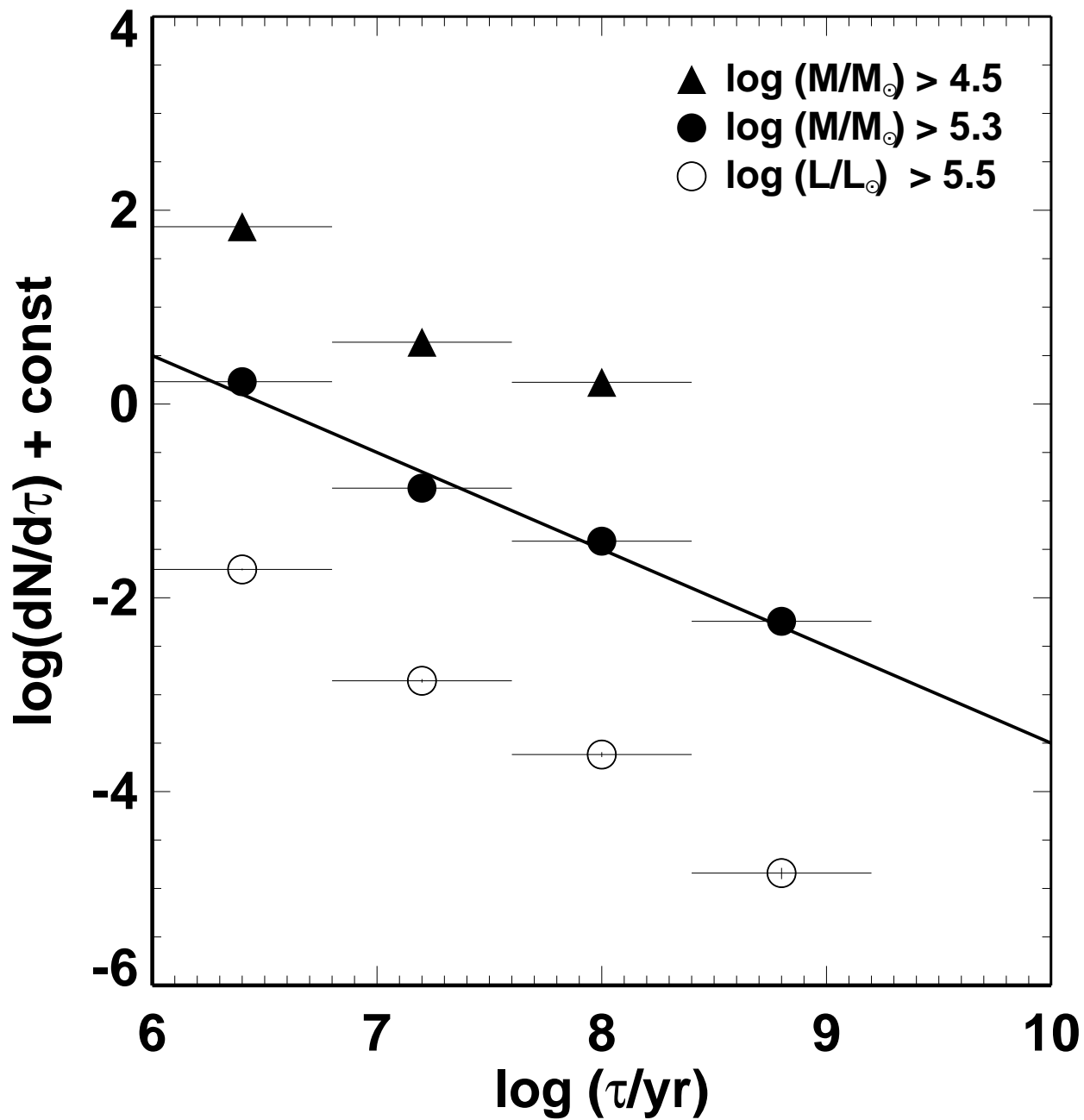}
\caption{Age distribution of star clusters in the Antennae
galaxies with different selection criteria.
The open symbols pertain to a luminosity-limited sample with
$L_V > 3 \times 10^5$~$L_{\odot}$, the approximate upper limit 
for stellar contamination, while the filled symbols pertain to
mass-limited samples with $M > 3\times10^4$~$M_{\odot}$ and
$M > 2\times10^5$~$M_{\odot}$.
The diagonal line represents the power law $dN/d\tau \propto
\tau^{-1}$, equivalent to $dN/d{\log\tau} = {\rm const}$.}
\end{figure}

\end{document}